\documentclass[preprint,amsmath,amssymb,prb,floatfix,superscriptaddress]{revtex4}

\usepackage[dvips]{graphicx}
\usepackage{dcolumn}
\usepackage{bm}
\usepackage{mathrsfs}

\begin{document}
\title{Spin-dependent polaron formation dynamics in Eu$_{0.75}$Y$_{0.25}$MnO$_3$ probed by femtosecond pump-probe spectroscopy}
\author{D. Talbayev}
 \email{dtalbayev@gmail.com}
 \affiliation{Department of Physics and Engineering Physics, Tulane University, 6400 Freret St., New Orleans, LA 70118, USA}
\author{Jinho Lee}
 \affiliation{Center for Integrated Nanotechnologies, MS K771, Los Alamos National Laboratory, Los Alamos, NM 87545, USA}
\author{S.A. Trugman}
 \affiliation{Center for Integrated Nanotechnologies, MS K771, Los Alamos National Laboratory, Los Alamos, NM 87545, USA}
\author{C.L. Zhang}
 \affiliation{Rutgers Center for Emergent Materials and Department of Physics and Astronomy, Rutgers University, Piscataway, NJ, 08854, USA}
\author{S.-W. Cheong} 
 \affiliation{Rutgers Center for Emergent Materials and Department of Physics and Astronomy, Rutgers University, Piscataway, NJ, 08854, USA}
\author{R.D. Averitt} 
 \affiliation{Department of Physics, Boston University, Boston, MA, 02215, USA}
 \affiliation{Department of Physics, UCSD, La Jolla, CA, 92093, USA}
\author{A. J. Taylor} 
 \affiliation{Center for Integrated Nanotechnologies, MS K771, Los Alamos National Laboratory, Los Alamos, NM 87545, USA}
\author{R. P. Prasankumar} 
 \affiliation{Center for Integrated Nanotechnologies, MS K771, Los Alamos National Laboratory, Los Alamos, NM 87545, USA}
\date{\today}

\newcommand{\cm}{\:\mathrm{cm}^{-1}}
\newcommand{\T}{\:\mathrm{T}}
\newcommand{\mc}{\:\mu\mathrm{m}}
\newcommand{\ve}{\varepsilon}
\newcommand{\dg}{^\mathtt{o}}
\newcommand{\ey}{Eu$_{0.75}$Y$_{0.25}$MnO$_3$}

\begin{abstract}
We present a femtosecond optical pump-probe study of the multiferroic manganite \ey.  The optical response of the material at pump energies of 1.55 and 3.1 eV is dominated by the $d$-$d$ and $p$-$d$ transitions of the Mn$^{3+}$ ions. The relaxation of photoexcited electrons includes the relaxation of the Jahn-Teller distortion and polaron trapping at Mn$^{2+}$ and Mn$^{4+}$ sites.  Ultrafast switching of superexchange interactions due to modulated $e_g$ orbital occupancy creates a localized spin excitation, which then decays on a time scale of tens of picoseconds at low temperatures.  The localized spin state decay appears as a tremendous increase in the amplitude of the photoinduced reflectance, due to the strong coupling of optical transitions to the spin-spin correlations in the crystalline $a$-$b$ plane.
\end{abstract}

\maketitle

\section{Introduction}
Strongly correlated electrons pose many important and difficult questions in modern condensed matter physics.  Correlated electron materials exhibit a wide range of physical properties resulting from the delicate balance between competing interactions that often involve several degrees of freedom, such as charge, spin, lattice, and orbital\cite{kotliar:53}.  In recent years, optical pump-probe spectroscopy has emerged as a promising spectroscopic tool for interrogating competing many-body interactions in the time domain\cite{basov:471,averitt:1357}.  This method relies on a femtosecond optical excitation pulse (the pump) that leaves the material in a nonequilibrium state; the subsequent electronic relaxation dynamics is recorded by another 'probe' pulse as the pump-induced change in the material's optical properties, e.g., the change $\Delta R$ in its reflectance\cite{averitt:1357}.  The electronic relaxation dynamics are governed by the many-body interactions that often act on different time scales\cite{averitt:1357,averitt:017401,talbayev:227002,talbayev:340}, which allows us to isolate competing interactions.

In this article, we focus on the Mott insulator \ey.  This compound is a relative of the prototypical Mott insulator LaMnO$_3$, the parent compound of the colossal magnetoresistance manganites\cite{salamon:583}.  In the insulating manganites with a perovskite structure, the magnetic Mn$^{3+}$ ions occupy the centers of the oxygen octahedra that form a corner-sharing network.  The compounds of this family exhibit a GdFeO$_3$-type distortion, in which the oxygen octahedra rotate about the pseudocubic [111] direction.  The neighboring octahedra rotate in the opposite direction. The Mn$^{3+}$ ions are found in the $t_{2g}^{3}e_g^1$ configuration, which makes their 3d orbital Jahn-Teller active\cite{salamon:583} and leads to orbital ordering and the cooperative Jahn-Teller distortion that sets in above 1000 K\cite{hemberger:035118}.  The smaller size of the Eu and Y ions relative to the La ion leads to a much stronger tilting of the oxygen octahedra and a much smaller Mn-O-Mn bond angle\cite{hemberger:035118,kimura:060403}, which results in the increased importance of the magnetic next-nearest-neighbor (NNN) superexchange interaction.  The magnetic structure of \ey\: is governed by the competition between the antiferromagnetic (AFM) NNN superexchange and the ferromagnetic nearest-neighbor (NN) superexchange  within the $ab$ plane.  The exchange interaction along the $c$ axis is AFM.  The frustrated nature of the magnetic exchange within the $ab$ plane leads to a low AFM transition temperature $T_N=47$ K and the long-wavelength modulation of the magnetic structure in the $ab$ plane\cite{hemberger:035118,kimura:060403}.  A remarkable property of \ey\: is the development of ferroelectricity\cite{hemberger:035118} at $T_{FE}=30$ K with ferroelectric polarization $\bm{P}$ in the $ac$ plane and $\bm{P_a}>>\bm{P_c}$.  This is believed to be driven by the formation of a magnetic cycloid\cite{hemberger:035118,mostovoy:067601}, similar to the observations in multiferroic TbMnO$_3$\cite{kimura:55,kenzelmann:087206}.  A new kind of magnetoelectric low-energy excitation was found in \ey, the electromagnon, which is described as a mixed magnon excitation that couples to the electric field of the light wave\cite{aguilar:060404, pimenov:014438, takahashi:214431, kubacka:1333}.

The present work is motivated by uncovering the interactions that govern the relaxation of photoexcited electrons in the magnetic and ferroelectric states of \ey, especially the effects of magnetoelectricity.  Over the past 15-20 years, time-resolved studies of magnetically ordered solids\cite{koopmans:844,talbayev:014417,kirilyuk:2731,zhao:207205} have grown steadily and matured into the field of femtomagnetism, which explores the fundamental aspects of the interactions between magnetic matter and light and the possibility of ultrafast control of the magnetic state.  On the other hand, optical time-resolved studies of multiferroics have been relatively rare in the past decade\cite{lim:4800, talbayev:097603, jang:031914, sheu:242904, qi:122904, handayani:116007, pohl:195112,lee:2654, kubacka:1333}.  We present an optical pump-probe study of the multiferroic \ey\: using femtosecond pump and probe pulses of 1.55 and 3.1 eV photon energies, in which we recorded the temporal evolution of the normalized pump-induced change in reflectance $\Delta R/R$.  We find a qualitatively similar response to photoexcitation with either wavelength.  At all temperatures, the $\Delta R/R$ decays on a long timescale of several nanoseconds, which we attribute to the recombination of photoinduced electrons and holes.

\begin{figure}[ht]
\begin{center}
\includegraphics[width=3in]{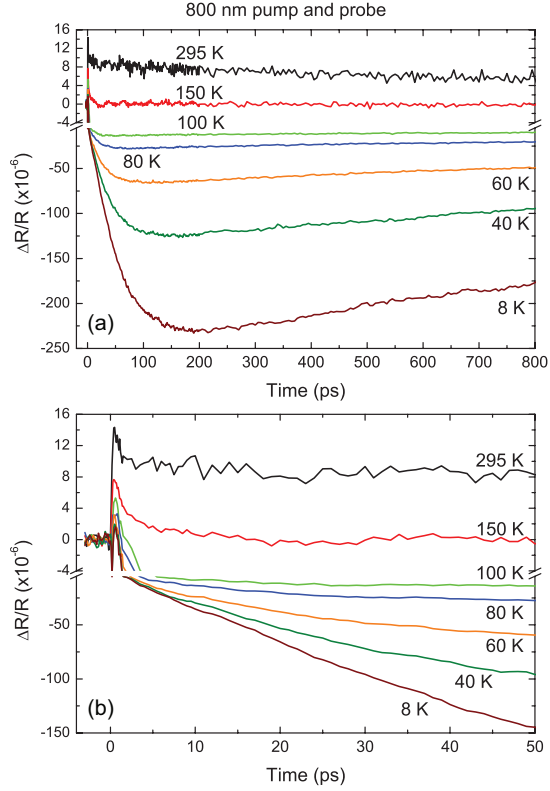}
\caption{\label{fig:pp800}(Color online) (a) Pump-probe reflectance spectra of the Eu$_{0.75}$Y$_{0.25}$MnO$_3$ crystal at various temperatures with 1.55 eV photon energy. (b) Same spectra as in (a), but zoomed in on the first 50 ps of photoinduced response.}
\end{center}
\end{figure}

The most remarkable feature of the pump-probe response is the development of a slow large-amplitude rise in the $\Delta R/R$ signal at low temperatures, with a maximum rise time of $\sim55$ ps (Fig.~\ref{fig:pp800}).  The feature starts to develop at temperatures above $T_N$ and reaches its full strength at the lowest temperatures.  We attribute this feature to the relaxation of the localized magnetic excitation that is created when a Mn$^{3+}$ ion is optically excited to Mn$^{2+}$ with an extra electron in the $e_g$ orbital.  This idea is analogous to lattice polaron formation at the Mn$^{2+}$ site, which happens as the doubly occupied $e_g$ orbital loses its Jahn-Teller activity.  This makes the Jahn-Teller distortion on the newly born Mn$^{2+}$ ion energetically unfavorable; the relaxation of the Jahn-Teller lattice distortion happens within $\sim1-5$ ps due to the strong electron-phonon coupling\cite{averitt:017401}.  This polaronic character of optical excitations in Jahn-Teller manganites is well established\cite{allen:4828,quijada:16093}.  The formation of the polaron can be viewed as the relaxation of the localized lattice excitation at the Mn$^{2+}$ site.  In a similar fashion, the Mn$^{2+}$ site is born with a localized magnetic excitation, because the magnetic exchange interactions between Mn$^{3+}$-Mn$^{2+}$ and Mn$^{3+}$-Mn$^{3+}$ are not the same\cite{salamon:583}.  The relaxation of this localized magnetic excitation happens via the emission of spin waves and takes much longer than the Jahn-Teller polaron relaxation.  

The long rise dynamics of the photoinduced response is a feature of the magnetically ordered state, but it emerges at $\sim100$ K, which is much higher than $T_N$.  We explain this by noting that the short-range spin correlations in magnetically frustrated systems persist to much higher temperatures compared to the long-range ordering temperature.  In \ey, the signatures of short-range spin correlations have been reported up to $\sim100$ K in the behavior of electromagnon and vibrational lattice excitations\cite{aguilar:060404,issing:024304}.  We find no significant changes in the pump-probe response at the temperature of the ferroelectric phase transition, $T_{FE}=30$ K, and interpret this as the negligible effect of magnetoelectric coupling in our measurement.  Our main finding is the magnetic relaxation triggered by the ultrafast switching of superexchange on the photoinjected Mn$^{2+}$ sites.  This process is generic, but very few works are found in the literature that report similar phenomena\cite{zhao:207205, wall:097402, handayani:116007}.  The corresponding magnitude of $\Delta R/R$ and the relaxation time depend strongly on the magnetic state of the material. At the lowest temperatures, this spin-dependent process dominates the photoinduced response.  The observed ultrafast modulation of the exchange interactions must be considered as a crucial ingredient in ultrafast control of magnetism and in the studies of photoinduced phase transitions.

\section{Experimental details}
Our optical pump-probe study of Eu$_{0.75}$Y$_{0.25}$MnO$_3$ was carried out on a single crystal grown using the flux method. We used the [001] oriented mechanically polished surface of the specimen, the $a$-$b$ plane. The pump-probe measurement employs pairs of laser pulses of 70 fs duration generated by a Ti:Sapphire regenerative amplifier (Coherent RegA) operating at a 250 kHz repetition rate. The first pulse, the pump, puts the material in a nonequilibrium excited state.  The second pulse, the probe, interrogates the photoinduced change in the material's reflectance. In all our measurements, the probe beam power was at least ten times lower than the power of the pump beam. We used lock-in detection to measure the pump-induced change in the reflected probe power and recorded the temporal evolution of the relative change in reflectance $\Delta R/R$.  In all measurements, the wavelength of the pump and the probe pulses was the same, corresponding to either 1.55 eV or 3.1 eV photon energy. The sample was mounted on the cold finger of a helium flow cryostat, which allowed control of the sample temperature in the 5 - 300 K range. 

\begin{figure}[ht]
\begin{center}
\includegraphics[width=3in]{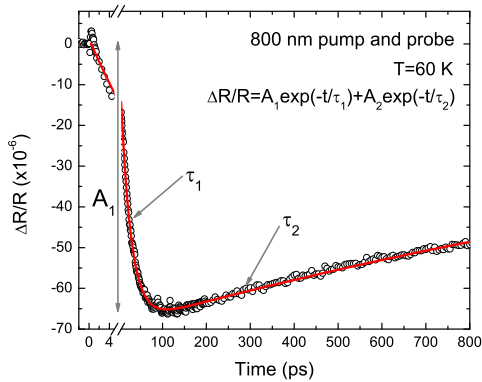}
\caption{\label{fig:g60k4mw}(Color online) Open circles - photoinduced change in reflectance at $T=60$ K at 1.55 eV. Solid line - fit to the data using the double exponential function in Eq.~(\ref{eq:doubleexpdecay}). Arrows and labels explain the meaning of the parameters in Eq.~(\ref{eq:doubleexpdecay}).  The long-rise dynamics are parametrized by the amplitude $A_1$ and the rise time $\tau_1$.}
\end{center}
\end{figure}

\begin{figure}[ht]
\begin{center}
\includegraphics[width=3in]{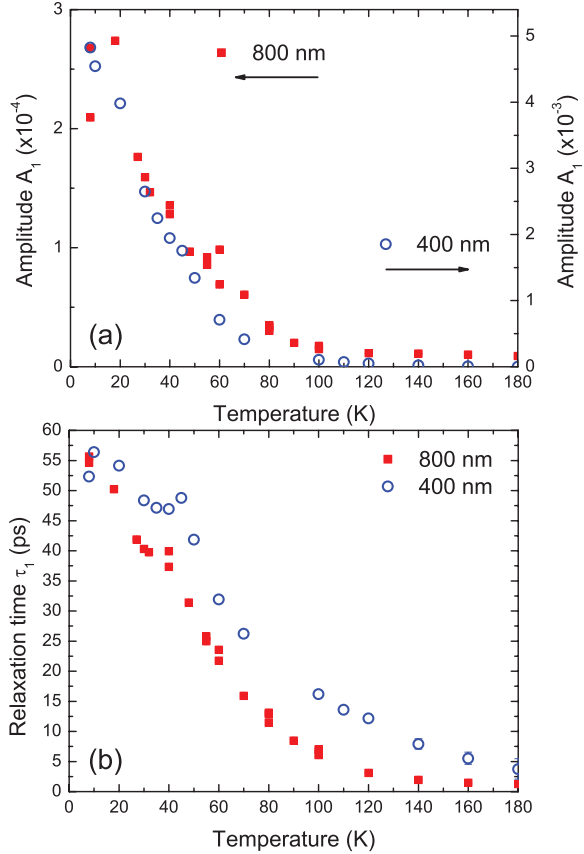}
\caption{\label{fig:a1tau1}(Color online) (a) Temperature dependence of the absolute value of the fitting parameter $A_1$. Solid symbols - the 1.55 eV measurement, left vertical axis. Open symbols - the 3.1 eV measurement, right vertical axis. (b) Temperature dependence of the rise-time of $\Delta R/R$ (the fitting parameter $\tau_1$). Solid symbols - the 1.55 eV  measurement. Open symbols - the 3.1 eV measurement.}
\end{center}
\end{figure}

Figure~\ref{fig:pp800} shows $\Delta R/R$ at 1.55 eV taken with a pump fluence of 50 $\mu J/cm^2$.  This fluence was chosen to be sufficiently low to keep the magnitude of $\Delta R/R$ proportional to the pump power and to be sufficiently high to provide a satisfactory signal-to-noise ratio.  Lowering the pump fluence to 5 $\mu J/cm^2$ led to a proportional drop in the amplitude of $\Delta R/R$ but did not change the observed time dynamics.  At high temperatures, the reflectance increases abruptly (in less than 0.5 ps) upon the arrival of the pump pulse. This jump in reflectance lasts for several nanoseconds and is due to the presence of photoinduced electrons and holes in the material. The positive reflectance jump becomes weaker with decreasing temperature, and $\Delta R/R$ develops a negative component below $\sim 150$ K. The time dependence of $\Delta R/R$ at low temperatures is well described by a double exponential function of the form
\begin{equation}
\Delta R/R=A_1\exp(-t/\tau_1) + A_2\exp(-t/\tau_2),
\label{eq:doubleexpdecay}
\end{equation}
where the first exponential term describes the rise of the negative response in $\Delta R/R$ (Fig.~\ref{fig:g60k4mw}). The second term describes the decay of the photoinduced reflectance within a few nanoseconds.  For simplicity, we did not include the convolution with the pump and probe autocorrelation function in Eq. (\ref{eq:doubleexpdecay}), as the dynamics of interest happens on a timescale much longer than the pulse duration.  From here on, we refer to the time constant $\tau_1$ as the rise time of the photoinduced reflectance. The most striking feature of our pump-probe data is the rapid increase in the magnitude of the negative $\Delta R/R$ component as the temperature is lowered below $100$ K. This increase can be quantified by extracting the amplitude $A_1$ from the least-square fits to the data using Eq.~(\ref{eq:doubleexpdecay}). Figure~\ref{fig:a1tau1} shows the temperature dependence of the parameters $A_1$ and $\tau_1$, both of which start rising steeply somewhere between 100 and 150 K as the temperature is lowered.  Figure~\ref{fig:a1tau1} also shows $A_1$ and $\tau_1$ extracted from the 3.1 eV pump-probe data of Fig.~\ref{fig:pp400}.  The amplitude $A_1$ at 3.1 eV is about ten times higher than at 1.55 eV and is positive, but their temperature dependences are very similar.  The behavior of the rise time $\tau_1$ in the 1.55 eV and 3.1 eV measurements is qualitatively very similar, although $\tau_1$ at 3.1 eV is longer at most temperatures.

\section{Discussion of the results}
\begin{figure}[ht]
\begin{center}
\includegraphics[width=3in]{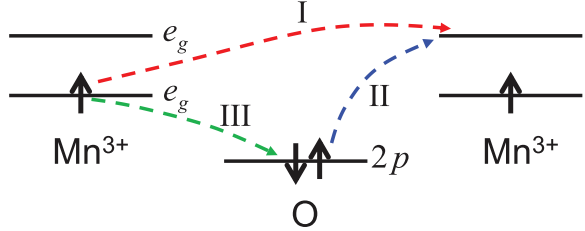}
\caption{\label{fig:transitions}(Color online) Optical transitions induced by the 1.55 eV and 3.1 eV pump.  Absorption of a 1.55 eV photon corresponds to transition I, which creates a pair of Mn$^{2+}$ and Mn$^{4+}$ ions.  Absorption of a 3.1 eV photon corresponds to transition II, which creates a Mn$^{2+}$ ion.  Transition III takes place after transition II and creates a Mn$^{4+}$ ion.}
\end{center}
\end{figure}

To understand the photoinduced response of \ey, we first need to consider its equilibrium optical conductivity.  The measured optical conductivity of the stoichiometric rare-earth perovskite manganites\cite{kovaleva:147204,bastjan:193105,moskvin:035106} displays two broad absorption peaks centered at $\sim2$ eV and $\sim5$ eV.  The lower-energy peak is dominated by intersite Mn $d$-$d$ transitions (transition I in Fig.~\ref{fig:transitions}), while the higher energy peak is dominated by on-site O $2p$-Mn $3d$ ($p$-$d$)charge-transfer transitions (transition II in Fig.~\ref{fig:transitions}).  The optical gap possesses mixed $d$-$d$ and $p$-$d$ character\cite{moskvin:035106}.  At the 1.55 eV pump excitation energy, the optical transitions mostly have $d$-$d$ character, while at the 3.1 eV energy they mostly have $p$-$d$ character\cite{kovaleva:147204, bastjan:193105, moskvin:035106}.  In both cases, an extra electron is promoted to the upper level of the Jahn-Teller split $e_g$ orbital, forming the Mn$^{2+}$ ion.  The subsequent relaxation dynamics of this excited state is recorded as the photoinduced reflectance $\Delta R/R$.  

\begin{figure}[ht]
\begin{center}
\includegraphics[width=3in]{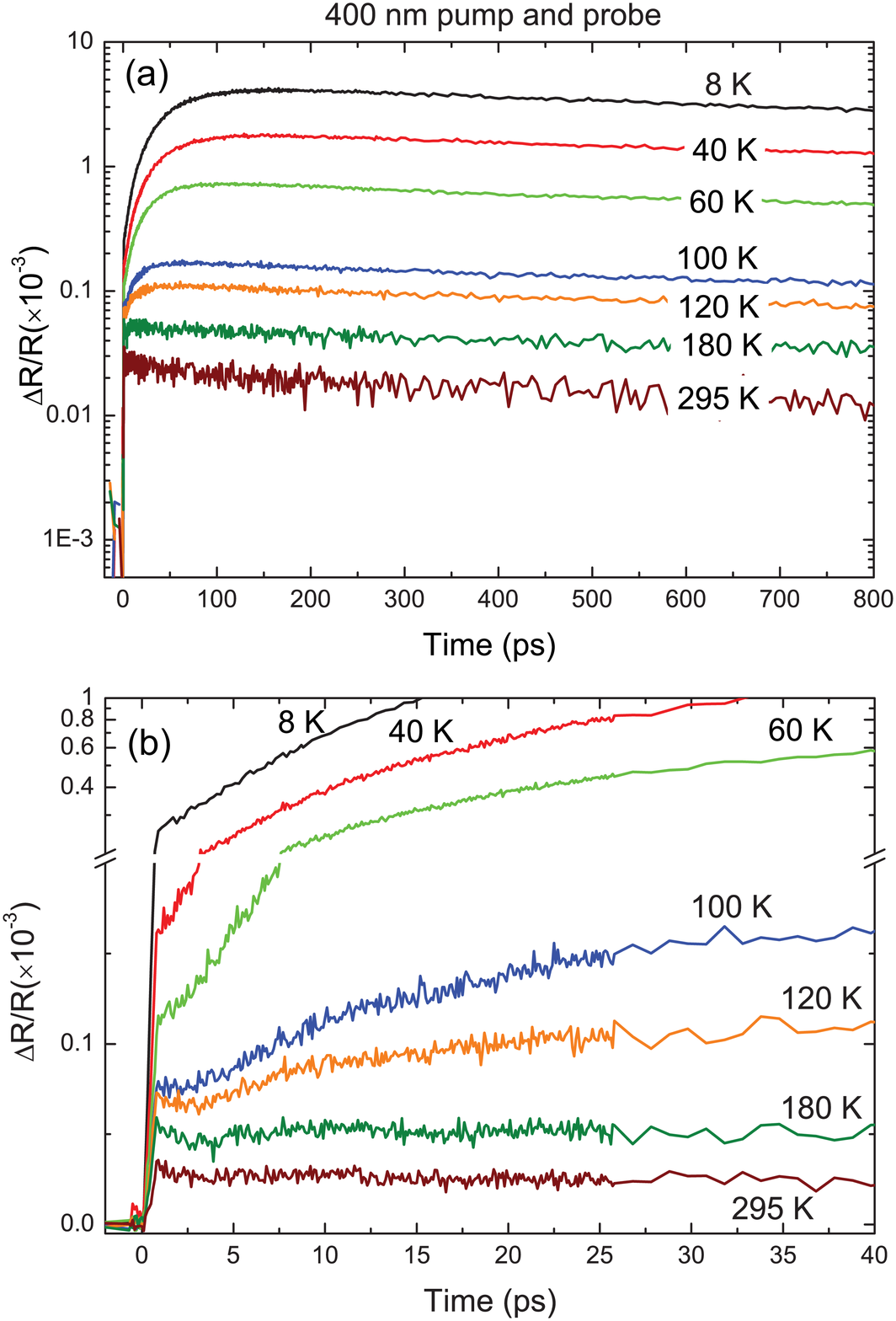}
\caption{\label{fig:pp400}(Color online) (a) Pump-probe reflectance spectra of the Eu$_{0.75}$Y$_{0.25}$MnO$_3$ crystal at various temperatures and the 3.1eV pump and probe photon energy. (b) Same spectra as in (a), but zoomed in on the first 40 ps of the photoinduced response.}
\end{center}
\end{figure}

The Mn$^{2+}$ ion is not Jahn-Teller active, but it is born with the Jahn-Teller distortion of its MnO$_6$ octahedron, which amounts to the displacive excitation of the localized Jahn-Teller phonon.  This mixed electron-phonon excited state quickly relaxes via electron-phonon and phonon-phonon couplings.  We attribute the short relaxation ($\leq5$ ps) clearly visible at high temperatures in the time-resolved spectra of Figs.~\ref{fig:pp800} and~\ref{fig:pp400} to the relaxation of the local Jahn-Teller distortion.  This relaxation is present in the spectra at all temperatures, but at low temperatures it is masked by the strong long-rise component of $\Delta R/R$ that we described by the rise time $\tau_1$ in the previous section.  At the end of the local Jahn-Teller relaxation, the extra $e_g$ electron finds itself trapped on the Mn$^{2+}$ site, as its hopping to the neighboring sites now costs half the energy of the Mn$^{3+}$ Jahn-Teller splitting.  Both theoretical and experimental evidence favors this self-trapped small polaron description of photodoped electrons in manganites\cite{allen:4828, quijada:16093, prasankumar:020402, wu:043901, ren:113013}. 

We now turn to the strongly temperature-dependent rise dynamics observed in Figs.~\ref{fig:pp800} and~\ref{fig:pp400} and described by the first term in Eq. (\ref{eq:doubleexpdecay}).  This long-rise dynamics dominates the photoinduced reflectance at low temperatures.  The contribution of this process to $\Delta R/R$ has opposite signs in the 1.55 eV (Fig.~\ref{fig:pp800}) and the 3.1 eV (Fig.~\ref{fig:pp400}) data.  Nonetheless, the two responses share the same origin.  This is corroborated by the similar temperature dependence of $\tau_1$ and $A_1$ in the two sets of data (Fig.~\ref{fig:a1tau1}).  The identical origin of the rise of $\Delta R/R$ is also consistent with the nature of the excited state, which is a self-trapped small polaron for both measurements.  It is not surprising that in both measurements we observe the same relaxation.  $\Delta R/R$ is related to the photoinduced modulation of the real and imaginary parts of the dielectric function, $\Delta\epsilon_1(t)$ and $\Delta\epsilon_2(t)$, as\cite{basov:471} 
\begin{equation}
\frac{\Delta R}{R}(t)=\frac{\partial \ln(R)}{\partial \epsilon_1}\Delta\epsilon_1(t) + \frac{\partial \ln(R)}{\partial \epsilon_2}\Delta\epsilon_2(t),
\label{eq:dreps}
\end{equation}
which means that the sign of $\Delta R/R$ depends on the ultrafast redistribution of the spectral weight.  In the following, we show that the long-rise response is magnetic in origin and that its magnitude is controlled by the strength of the spin correlations in \ey.  The spin correlations also control the spectral weight redistribution and the dielectric function change $\Delta\epsilon$ with temperature.  For example, cooling LaMnO$_3$ from 300 K to 50 K causes positive changes $\Delta\epsilon_1>0$ and $\Delta\epsilon_2>0$ at 1.55 eV and negative changes $\Delta\epsilon_1<0$ and $\Delta\epsilon_2<0$ at 3.1 eV\cite{kovaleva:147204}.  The Ne\'el temperature of LaMnO$_3$ is 140 K.  Similarly, in our ultrafast measurement, the opposite sign of $\Delta R/R$ at 1.55 eV and 3.1 eV may be related to the spin-correlation-controlled dynamic redistribution of the spectral weight.

The low-temperature properties of \ey\: are dominated by two phenomena - the development of AFM long-range order at $T_N=47$ K and of ferroelectricity at $T_{FE}=30$ K.  The temperature dependence of $A_1$ and $\tau_1$ does not show any significant features at those transitions (Fig.~\ref{fig:a1tau1}); $A_1$ emerges smoothly at around 100 K and continues its growth down to the lowest temperatures.  In the literature, we find two instances of a change in the physical properties of \ey\: that develops at 100 K and gets stronger with decreasing temperature.  In the study of electromagnons by Vald\'es Aguilar et al.\cite{aguilar:060404}, the far-infrared spectral weight in the optical conductivity begins to grow just above $100$ K and continues to increase through both phase transitions as the temperature is lowered.  In the Raman study of phonons by Issing et al.\cite{issing:024304}, several phonon frequencies start to soften at $\sim100$ K and continue their softening down to the lowest temperature.  Both the far-infrared spectral weight\cite{aguilar:060404} and the phonon softening\cite{issing:024304} exhibit temperature dependence that is very similar to that of $A_1$ (Fig.~\ref{fig:a1tau1}).  In both cases, the temperature changes are caused by the strong spin-phonon coupling characteristic of multiferroics, and the magnitude of the change is governed by the strength of the spin correlations.  Based on the similar temperature dependence of the amplitude $A_1$, we propose that the low-temperature pump-probe response is also controlled by the coupling to spin correlations.  The presence of such correlations well above $T_N$ in \ey\: is expected because of the frustrated magnetic interactions\cite{kimura:060403}.  

\begin{figure}[ht]
\begin{center}
\includegraphics[width=3in]{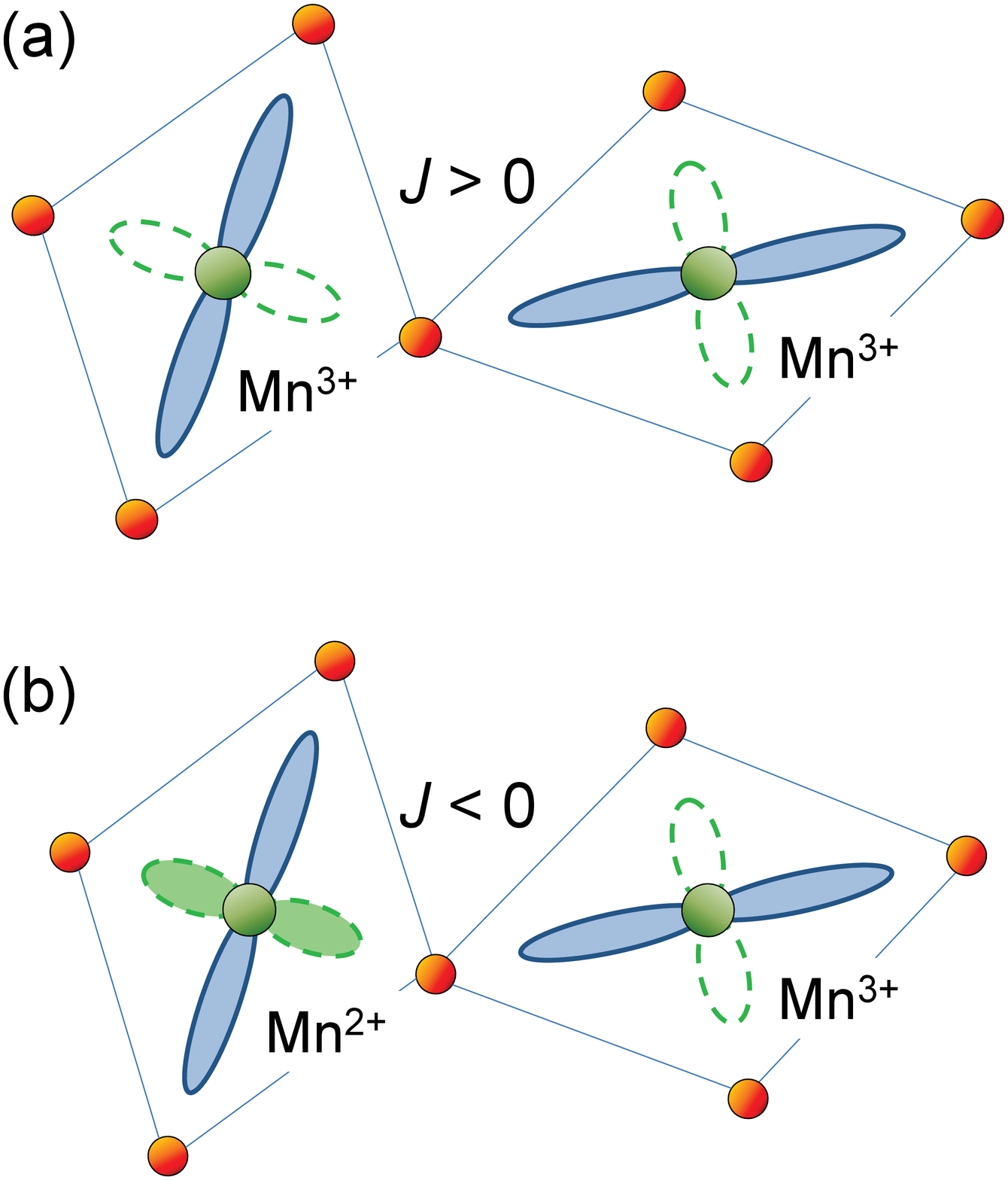}
\caption{\label{fig:orbitals}(Color online) Sketch of the neighboring Mn ions and their oxygen environment in the $a$-$b$ plane of \ey.  The spheres with lobes represent the Mn ions and their $e_g$ orbitals.  (a) The ground state with two Mn$^{3+}$ ions.  The filled lobes show the half-filled $3z^2-r^2$ orbitals of the $e_g$ manifold.  The dashed empty lobes show the empty $x^2-y^2$ orbitals.  The exchange-mediating oxygen $2p$ orbital (not shown) overlaps an empty and a half-filled $e_g$ orbital, leading to ferromagnetic exchange with $J>0$.  (b) The photoexcited state with Mn$^{2+}$ and Mn$^{3+}$ ions.  The $x^2-y^2$ orbital is now half-filled, leading to antiferromagnetic exchange with $J<0$.}
\end{center}
\end{figure}

We now describe the mechanism by which the spin correlations result in the long rise dynamics found in our pump-probe spectra.  The photoexcitation of a second electron into the $e_g$ manifold switches the sign of the superexchange interaction of the Mn ion with some of its nearest neighbors.  In the $a$-$b$ plane, the orbital order makes the Mn$^{3+}$-Mn$^{3+}$ superexchange ferromagnetic (positive exchange constant $J$) according to the Goodenough-Kanamori rules\cite{goodenough:564}, as the mediating oxygen $p$ orbital overlaps a half-filled and an empty $e_g$ orbital (Fig.~\ref{fig:orbitals}).  Photoexciting an electron to the empty $e_g$ orbital creates the Mn$^{2+}$ ion and changes the sign of the constant $J$, as the oxygen $p$ orbital now overlaps two half-filled $e_g$ orbitals.  This ultrafast modulation of the exchange coupling between nearest-neighbor Mn ions creates a localized spin excitation on the Mn$^{2+}$ site, which is analogous to the localized polaron excitation that we considered above.  An equivalent picture of carrier (hole) trapping and local spin excitation can be constructed for the Mn$^{4+}$ sites, as they are also not Jahn-Teller active and the exchange coupling of the Mn$^{4+}$ ion with some of its Mn$^{3+}$ neighbors has an opposite sign compared to the Mn$^{3+}$-Mn$^{3+}$ superexchange.

We attribute the emergence of the long rise dynamics to the decay of the localized spin excitations via the emission of spin waves and spin-lattice coupling.  The rise time $\tau_1$ reflects the combined lifetime of the localized spin excitations on Mn$^{2+}$ and Mn$^{4+}$ sites.  In this picture, the increase of amplitude $A_1$ with lower temperature (Fig.~\ref{fig:a1tau1}) is linked to the strengthening of the spin correlations in the material: the stronger correlations result in a larger amplitude of the localized spin excitation.  The stronger spin correlations also make our pump-probe measurement more sensitive to the local spin structure on Mn sites, as they control the strength of the optical $d$-$d$ and $p$-$d$ transitions\cite{kovaleva:147204}.  We interpret the increase in the rise time $\tau_1$ (Fig.~\ref{fig:a1tau1}) as the increasing lifetime of the localized spin excitation with lower temperature, which should also grow with the strength of the spin correlations.  This is equivalent to the narrowing of electron spin resonance lines in the magnetically ordered state of manganites\cite{mihaly:024414}.

We interpret the observed rise time $\tau_1$ in the 1.55 eV and 3.1 eV data as the signature of the same process - the decay of localized spin excitations.  The measured values of $\tau_1$ are somewhat different for the two photon energies (Fig.~\ref{fig:a1tau1}).  We would like to discuss a possible origin of that difference.  The absorption of a 1.55 eV photon creates a pair of Mn$^{2+}$ and Mn$^{4+}$ ions (transition I in Fig.~\ref{fig:transitions}).  The absorption of a 3.1 eV photon also creates such a pair, but in a two-step process.  In the first step, the transition II in Fig.~\ref{fig:transitions} takes place, leaves a hole on the oxygen ion, and creates a Mn$^{2+}$ ion.  In the second step, the hole on the oxygen is filled and transferred to a neighboring Mn ion, turning it into Mn$^{4+}$ (transition III in Fig.~\ref{fig:transitions}).  Thus, the creation of the Mn$^{4+}$ ion may take longer when \ey\: is excited with the 3.1 eV pump, which may lead to the longer rise time $\tau_1$ observed at this photon energy.

Handayani et al.\cite{handayani:116007} performed an optical pump-probe study of multiferroic manganite TbMnO$_3$, which is very similar to \ey\: in many respects, e.g., in its electronic structure, the orbital ordered state, and the magnetic and multiferroic properties.  They found a very similar pump-probe phenomenology in TbMnO$_3$, i.e., the emergence of a long rise in the photoinduced $\Delta R/R$ that dominates the response at low temperature.  They also assigned a magnetic origin to this long rise component, but their microscopic interpretation is different from ours.  They interpret it as the loss of kinetic energy by the photoexcited electron via the emission of spin waves (or magnon assisted hopping of the electron).  The rise time of $\Delta R/R$ is then the amount of time it takes the photoinjected electron to stop hopping from site to site\cite{handayani:116007}.  In our view, this picture does not account for the electron-phonon coupling that leads to the trapping of the polaron that we discussed earlier.

\section{Summary}
We have presented an optical pump-probe study of the multiferroic manganite \ey\:.  Photoexcitation promotes an extra electron to the Mn $e_g$ manifold, which then forms a self-trapped polaronic state due to the local relaxation of the Jahn-Teller distortion on a picosecond time scale.  At low temperatures, the process that dominates the $\Delta R/R$ response is the decay of the localized spin excitation created on the polaron site.  The photoexcitation of an extra electron to the $e_g$ manifold triggers the ultrafast switching of the superexchange interaction of the Mn$^{2+}$ and Mn$^{4+}$ ions with their neighbors according to the Goodenough-Kanamori rules\cite{goodenough:564}, creating the localized spin excitation.  The decay of the localized spin excitation appears in our spectra as the very large amplitude ($A_1$) modulation of $\Delta R/R$, which is controlled by the sensitivity of the $d$-$d$ and $p$-$d$ optical transitions to the spin correlations in \ey.  The ultrafast modulation of the exchange interaction is a very general process, but few similar observations have been reported\cite{wall:097402, zhao:207205, handayani:116007}.  This fundamental opto-magnetic process is important for the models of light-matter interactions used to describe femtomagnetic phenomena\cite{kirilyuk:2731}.

The work at Tulane was supported by the Louisiana Board of Regents through the Board of Regents Support Fund contract number LEQSF(2012-15)-RD-A-23.  The work at Los Alamos National Laboratory was supported by the LDRD program and by the Center for Integrated Nanotechnologies.  The work at Rutgers University was supported by DOE: DE-FG02-07ER46382.

\newpage

\end{document}